\pdfoutput=1

\documentclass{amsart}

\usepackage{amsmath,amssymb}
\usepackage{amsthm, amsrefs}
\usepackage{latexsym}
\usepackage{indentfirst}
\usepackage{mathrsfs}
\usepackage{placeins}
\usepackage{graphicx}

\numberwithin{equation}{section}

\theoremstyle{plain}

\theoremstyle{definition}

\newtheorem{assump}{Assumption}

\theoremstyle{remark}
\newtheorem*{remark}{Remark}

\newcommand{\etal}{\textit{et al}\;}
\newcommand{\ie}{\textit{i.e.}\;}
\newcommand{\eg}{\textit{e.g.}\;}

\newcommand{\ud}{\,\mathrm{d}}

\newcommand{\TT}{\mathrm{T}}

\newcommand{\Or}{\mathcal{O}}

\newcommand{\bd}[1]{\boldsymbol{#1}}

\newcommand{\veps}{\varepsilon}

\newcommand{\abs}[1]{\lvert#1\rvert}

\newcommand {\f}{\frac}
\newcommand {\p}{\partial}

\newcommand{\beq}{\begin{equation}}
\newcommand{\eeq}{\end{equation}}

\newcommand{\barint}{\kern3pt \raise3.4pt\hbox{\vrule height.6pt
    width7pt} \kern-10pt \int}

\begin{document}

\title[A mean-field model for chemotaxis]{A pathway-based mean-field model for \emph{E. coli}
chemotaxis: Mathematical derivation and Keller-Segel limit}

\author{Guangwei Si}
\address{Center for Quantitative Biology \\
 Peking University \\
Beijing, China, 100871\\
email: gwsi@pku.edu.cn }

\author{Min Tang}
\address{Institute of Natural Sciences, Department of mathematics and MOE-LSC \\
Shanghai Jiao Tong University, 200240, Shanghai, China\\
email:tangmin@sjtu.edu.cn }

\author{Xu Yang}
\address{Department of Mathematics \\
  University of California \\
  Santa Barbara, CA 93106 \\
  email: xuyang@math.ucsb.edu }

\date{\today}

\thanks{G.S. was partially supported by NSF of China
under Grants No. 11074009 and No. 10721463 and the MOST of China
under Grants No. 2009CB918500 and No. 2012AA02A702. M.T. was
partially supported by Shanghai Natural Science Foundation of
Shanghai under Grant No. 12ZR1445400. X.Y. was partially supported
by the startup funding of Department of Mathematics, University of
California, Santa Barbara. G.S. would like to thank Yuhai Tu for
valuable discussions and Tailin Wu for his early work on
simulation.}

\maketitle

\begin{abstract}

A pathway-based mean-field theory (PBMFT) was recently proposed for
\emph{E. coli} chemotaxis in [G. Si, T. Wu, Q. Quyang and Y. Tu,
{\it Phys. Rev. Lett.}, 109 (2012), 048101]. In this paper, we
derived a new moment system of PBMFT by using the moment closure
technique in kinetic theory under the assumption that the
methylation level is locally concentrated. The new system is
hyperbolic with linear convection terms. Under certain assumptions,
the new system can recover the original model. Especially the
assumption on the methylation difference made there can be
understood explicitly in this new moment system. We obtain the
Keller-Segel limit
 by taking into account the
different physical time scales of tumbling, adaptation and
the experimental observations. We also present numerical evidence to show the
quantitative agreement of the moment system with the individual
based \emph{E. coli} chemotaxis simulator.
\end{abstract}

\section{introduction}

The locomotion of \emph{Escherichia coli} (\emph{E. coli}) presents
a tumble-and-run pattern (\cite{BeBr:72}), which can be viewed as a
biased random walk process. In the presence of chemoeffector with
nonzero gradients, the suppression of direction change (tumble)
leads to chemotaxis toward the high concentration of
chemoattractants (\cite{Ad:66,Be:00}). Great efforts have been put
into understanding the chemotactic sensory system of \emph{E. coli}
(\cite{WaAr:04,TiPoMaGaAr:08,HaFaPa:08}). The chemotaxis signaling
pathway belongs to the class of two-component sensory system, which
consists of sensors and response regulators. The chemotaxis sensor
complex is composed of transmembrane chemo-receptors, the adaptor
protein CheW, and the histidine kinase CheA. The response regulator
CheY controls the tumbling frequency of the flagellar motor
(\cite{HaHa:83}). Adaptation is carried out by the two enzymes, CheR
and CheB, which control the kinase activity by modulating the
methylation level of receptors (\cite{WaAr:04}). Because of slow
adaptation process, receptor methylation level serves as the memory
of cells, and cells decide whether to run or tumble by comparing
receptor methylation level to local environments.

In the modeling literature, bacterial chemotaxis has been described
by the Keller-Segel (K-S) model at the population level
(\cite{KeSe:71}), where the drift velocity is given by empirical
functions of chemoeffector gradient. It has successfully explained
the chemotaxis phenomenon in slowly changing environments
(\cite{TiMAPoAr:08}), however fails to make good predictions in
rapidly changing ones (\cite{Zhuetal:12}) and the volcano effects
(\cite{SiMi:11,BrLeLi:07}). Besides that, the K-S model has also
been mathematically proved to present nonphysical blowups in high
dimension when initial total mass reaches the critical level
(\cite{BlDoPe:06,BlCaMa:08,BiCoDo:09}). In order to understand
bacterial behavior from the individual dynamics, kinetic models have
been also developed by considering the velocity-jump process
(\cite{Al:80,St:00,HiOt:00}), and the K-S model can be
systematically derived by taking the hydrodynamic limit of kinetic
models (\eg \cite{ChMaPeSc:04,FiLaPe:05}). All the above mentioned
models are phenomenological and do not take into account the
signaling transduction and adaptation process.

Nowadays, modern experimental technologies have been able to
quantitatively measure the dynamics of signaling pathways of
\emph{E. coli} (\cite{AlSuBaLe:99,ClSuLe:00,SoBe:02,ShTuBe:10}),
which has led to successful modeling of the pathway dynamics
(\cite{MeTu:03,Keetal:06,TuShBe:08}). These works make possible the
verification of predictive agent-based models that include the
intracellular signaling pathway dynamics. It is of great biological
interest to understand the molecular origins of chemotaxis behavior
of \emph{E. coli} by deriving population-level model based on the
underlying signaling pathway dynamics (\cite{ErOt:04,SiWuOuTu:12}).
Particularly in \cite{SiWuOuTu:12}, the authors developed a
pathway-based mean field theory (PBMFT) that incorporated the most
recent quantitatively measured signaling pathway, and explained a
counter-intuitional experimental observation which showed that in a
spatial-temporal fast-varying environment, there exists a phase
shift between the dynamics of ligand concentration and center of
mass of the cells \cite{Zhuetal:12}. Especially, when the
oscillating frequency is comparable to the adaptation rate of
\emph{E. coli}, the phase shift becomes significant. Apparently this
is a phenomenon that can not be explained by the K-S model.

In this paper, we study PBMFT for \emph{E. coli} chemotaxis from a
mathematical point of view. Specifically we derive a new moment
system of PBMFT using the moment closure technique in kinetic
theory. The new system is hyperbolic with linear convection terms.
Under certain assumptions, the derived moment system gets to the
original model in \cite{SiWuOuTu:12}, and especially the assumption
on the methylation difference made in \cite{SiWuOuTu:12} can be
understood explicitly in this new system. Taking into account the
different physical time scales of the tumbling, adaptation and
experimental observation, we connect the moment system to the K-S
model (in the parabolic scaling). The agreement of the moment system
with the signaling pathway-based \emph{E. coli} chemotaxis
agent-based simulator (SPECS \cite{JiOuTu:10}) will be provided
numerically in the environment of spatial-temporal varying ligand
concentration.

The rest of the paper is organized as follows. We introduce the
pathway-based kinetic model incorporating the intracellular
adaptation dynamics in Section~\ref{sec:model}. In
Section~\ref{sec:1dMFT}, assuming the methylation level is locally
concentrated, we are able to build the moment system by using the
moment closure technique in one dimension. Furthermore, the modeling
assumption will be justified both analytically and numerically.
Section~\ref{sec:KS} illustrates why K-S model is valid in the slow
varying environments. We also give the connection of the moment
system to the PBMFT model proposed in \cite{SiWuOuTu:12}, and
provide the quantitative agreement of the moment system with SPECS
numerically. Two-dimensional moment system is derived in
Section~\ref{sec:2dMFT}, and we make conclusive remarks in
Section~\ref{sec:conclusion}.

\section{Description of the kinetic model}\label{sec:model}
We shall start from the same kinetic model used in
\cite{SiWuOuTu:12}, which incorporates the most recent progresses on
the chemo-sensory system (\cite{TuShBe:08, ShTuBe:10}). The model is
a one-dimensional two-flux model given by
\begin{align}\label{eq:kineticmodel1}
\f{\p P^+}{\p t}&=-\f{\p(v_0 P^+)}{\p x}-\f{\p (f(a)P^+)}{\p m}-\f{z(m)}{2}(P^+-P^-),\\
\f{\p P^-}{\p t}&=\f{\p(v_0 P^-)}{\p x}-\f{\p (f(a)P^-)}{\p
m}+\f{z(m)}{2}(P^+-P^-).\label{eq:kineticmodel2}
\end{align}
In this model, each single cell of \emph{E. coli} moves either in
the ``$+$'' or ``$-$'' direction with a constant velocity $v_0$.
$P^\pm(t,x,m)$ is the probability density function for the cells
moving in the ``$\pm$'' direction, at time $t$, position $x$ and
methylation level $m$.

The intracellular adaptation dynamics is described by
\beq\label{eq:Fa} \frac{\ud m}{\ud t}=f(a)=k_R(1-a/a_0), \eeq where
the receptor activity $a(m,[L])$ depends on the intracellular
methylation level $m$ as well as the extracellular chemoattractant
concentration $[L]$, which is given by \beq\label{eq:am}
a=\bigl(1+\exp(NE)\bigr)^{-1}.\eeq According to the two-state model
in \cite{MeTu:03, Keetal:06}, the free energy is \beq
\label{eq:E}E=-\alpha(m-m_0)+f_0([L]), \qquad\mbox{with}\quad
f_0([L])=\ln\biggl(\f{1+[L]/K_I}{1+[L]/K_A}\biggr). \eeq
 In \eqref{eq:Fa}, $k_R$ is the
methylation rate, $a_0$ is the receptor preferred activity that
satisfies $f(a_0)=0$, $f'(a_0)<0$. $N$, $m_0$, $K_I$, $K_A$
represent the number of tightly coupled receptors, basic methylation
level, and dissociation constant for inactive receptors and active
receptors respectively.

We take the tumbling rate function $z(m,[L])$ in \cite{SiWuOuTu:12},
\beq\label{eq:Z} \quad z=z_0+\tau^{-1}(a/a_0)^H, \eeq where $z_0$,
$H$, $\tau$ represent the rotational diffusion, the Hill coefficient
of flagellar motor's response curve and the average run time
respectively. We refer the readers to \cite{SiWuOuTu:12} and the
references therein for the detailed physical meanings of these
parameters.

More generally, the kinetic model incorporating chemo-sensory system
is given as below,
\begin{equation}\label{eq:kinetic2d}
{\p_t}P=-\bd{v}\cdot\nabla_{\bd{x}} P-\p_m(f(a)P)+Q(P,z),
\end{equation}
where $P(t,\bd{x},\bd{v},m)$ is the probability density function of
bacteria at time $t$, position $\bd{x}$, moving at velocity $\bd{v}$
and methylation level $m$.

The tumbling term $Q(P,z)$ is
\begin{equation}\label{eq:Q}
Q(P,z)=\barint_\Omega
z(m,[L],\bd{v},\bd{v}')P(t,\bd{x},\bd{v}',m)\ud\bd{v}'
-\barint_\Omega
z(m,[L],\bd{v}',\bd{v})\ud\bd{v}'P(t,\bd{x},\bd{v},m),
\end{equation}
where $\Omega$ represents the velocity space and the integral
$$\barint=\frac{1}{\abs{\Omega}}\int_\Omega,\qquad\mbox{where }
\abs{\Omega}=\int_\Omega\ud\bd{v},$$ denotes the average over
$\Omega$. $z(m,[L],\bd{v},\bd{v}')$ is the tumbling frequency from
$\bd{v}'$ to $\bd{v}$, which is also related to the activity $a$ as
in \eqref{eq:Z}. The first term on right-hand side of \eqref{eq:Q}
is a gain term, and the second is a loss term.

\section{One-dimensional mean-field model}\label{sec:1dMFT}

In this section, we derive a new moment system of PBMFT from
\eqref{eq:kineticmodel1}-\eqref{eq:kineticmodel2} based on the the
assumption that the methylation level is locally concentrated. This
assumption will be justified by the numerical simulations using
SPECS and the formal analysis in the limit of $k_R \rightarrow
\infty$. To simplify notations, we denote $\int_0^{+\infty}$ by
$\int$ in the rest of this paper.

\subsection{Derivation of a new moment system of PBMFT}
Firstly, we define the macroscopic quantities, density, density
flux, momentum (on $m$) and momentum flux as follows,

\begin{align}
&\rho(x,t)=\int (P^+ + P^-)\ud m, \quad
J_\rho(x,t)=\int v_0(P^+ - P^-)\ud m;\label{eq:rhojrho}\\
& q(x,t)=\int m (P^+ + P^-) \ud m, \quad J_q(x,t)=\int v_0 m(P^+ -
P^-) \ud m. \label{eq:qjq}
\end{align}
The average methylation level $M(t,x)$ is defined as

\begin{equation}\label{eq:M}
 M=\frac{q}{\rho}.
\end{equation}
For simplicity, we also introduce the following notations
\begin{equation}\label{eq:PF}
\begin{aligned}
& Z=z\bigl(M(t,x)\bigr),
\qquad \f{\p Z}{\p m}=\f{\p z}{\p m}\Big\vert_{m=M}, \\
&F=f\Bigl(a\bigl(M(t,x)\bigr)\Bigr), \qquad \f{\p F}{\p m}=\f{\p
f}{\p m}\Big\vert_{m=M}.
\end{aligned}
\end{equation}

\begin{assump}\label{asp:closure} We need the following condition to close the
moment system,
\begin{equation*}
\f{\int (m-M)^2P^\pm \ud m}{\int P^\pm \ud m} \ll 1.
\end{equation*}
\end{assump}

\begin{remark}
Physically this assumption means, the distribution functions
$P^\pm$ is localized in $m$, and the variation of averaged
methylation is
small in both moving directions ``$\pm$''. 
\end{remark}

Integrating \eqref{eq:kineticmodel1}$+$\eqref{eq:kineticmodel2}
with respect to $m$ yields the equation for density \begin{equation*}
\frac{\partial \rho}{\partial t}+\frac{\partial J_\rho}{\partial
x}=0. \end{equation*} Integrating
$v_0\times$\eqref{eq:kineticmodel1}$-$
$v_0\times$\eqref{eq:kineticmodel2} with respect to $m$ produces
\begin{equation*}
\begin{aligned}
\f{\p J_\rho}{\p t}&=-v_0^2\f{\p \rho}{\p x}-v_0\int z(m)(P^+ - P^-) \ud m\\
&\approx -v_0^2\f{\p \rho}{\p x}-v_0\int \biggl(z(M)+\f{\p z}{\p
m}\Big\vert_{m=M}(m-M) \biggr)
(P^+ - P^-) \ud m \\
&=-v^2_0\f{\p \rho}{\p x}-ZJ_\rho-\f{\p Z}{\p m}(J_q-MJ_\rho),
\end{aligned}
\end{equation*}
where we have used Assumption~\ref{asp:closure} in the second step
and the notations in \eqref{eq:M}, \eqref{eq:PF} in the third
step.

Similarly, integrating $m\times$\eqref{eq:kineticmodel1}$+$
$m\times$\eqref{eq:kineticmodel2} with respect to $m$ gives
\begin{equation*}
\begin{aligned}
\f{\p q}{\p t}&=-\f{\p J_q}{\p x}+\int f(a)(P^+ + P^-) \ud m\\
&\approx -\f{\p J_q}{\p x}+\int \biggl(f(a)\vert_{m=M}+\f{\p
f}{\p m}\Big\vert_{m=M}(m-M) \biggr)(P^+ + P^-) \ud m \\
&=-\f{\p J_q}{\p x}+F\rho+\f{\p F}{\p m}(q-M\rho)\\
&=-\f{\p J_q}{\p x}+F\rho,
\end{aligned}
\end{equation*}
where we have used an integration by parts in the first step and
the definition of $M$ in \eqref{eq:M} in the last step.

Integrating $v_0m\times$\eqref{eq:kineticmodel1}$-$
$v_0m\times$\eqref{eq:kineticmodel2} with respect to $m$ yields
\begin{equation*}
\begin{aligned}
\f{\p J_q}{\p t}=&-v_0^2\f{\p q}{\p x}+v_0\int f(a)(P^+ - P^-) \ud
m - v_0\int z(m)m(P^+ - P^-) \ud m
\\
\approx&-v_0^2\f{\p q}{\p x}+v_0\int\biggl(f(a)\vert_{m=M}+\f{\p
f}{\p m}\Big\vert_{m=M}(m-M) \biggr)(P^+-P^-)\ud m\\
&-v_0\int\biggl((z(m)m)\vert_{m=M}+\f{\p (z(m)m)}{\p
m}\Big\vert_{m=M}(m-M) \biggr)(P^+-P^-)\ud m
\\
=&-v_0^2\f{\p q}{\p x}+FJ_\rho+\f{\p F}{\p
m}(J_q-MJ_\rho)-ZMJ_\rho
-\Big(\f{\p Z}{\p m}M+Z\Big)(J_q-MJ_\rho)\\
=& -v_0^2\f{\p q}{\p x}+FJ_\rho+\f{\p F}{\p m}(J_q-MJ_\rho)-ZJ_q
-\f{\p Z}{\p m}M(J_q-MJ_\rho),
\end{aligned}
\end{equation*}

where we have used Assumption~\ref{asp:closure} in the last step.

Altogether, we obtain a closed moment system for $\rho$, $J_{\rho}$,
$q$ and $J_q$
\begin{align}
\label{eq:density} \frac{\partial \rho}{\partial t}&
=-\frac{\partial
J_\rho}{\partial x}, \\
\label{eq:densityflux} \f{\p J_\rho}{\p t}&=-v^2_0\f{\p \rho}{\p
x}-ZJ_\rho-\f{\p Z}{\p m}(J_q-MJ_\rho), \\
\label{eq:momentum} \f{\p q}{\p t}&=-\f{\p J_q}{\p x}+F\rho, \\
\label{eq:momentumflux} \f{\p J_q}{\p t} &= -v_0^2\f{\p q}{\p
x}+FJ_\rho+\f{\p F}{\p m}(J_q-MJ_\rho)-ZJ_q -\f{\p Z}{\p
m}M(J_q-MJ_\rho).
\end{align}

\begin{remark}

The Taylor expansion in $m$ gives a systematical way of constructing
high order moment systems. Please see the Appendix for the
derivation of second-order moment system.
\end{remark}

\subsection{Numerical Justification of Assumption~\ref{asp:closure} by SPECS}



\begin{figure}[h]
\centering
\includegraphics[width=5.0in]{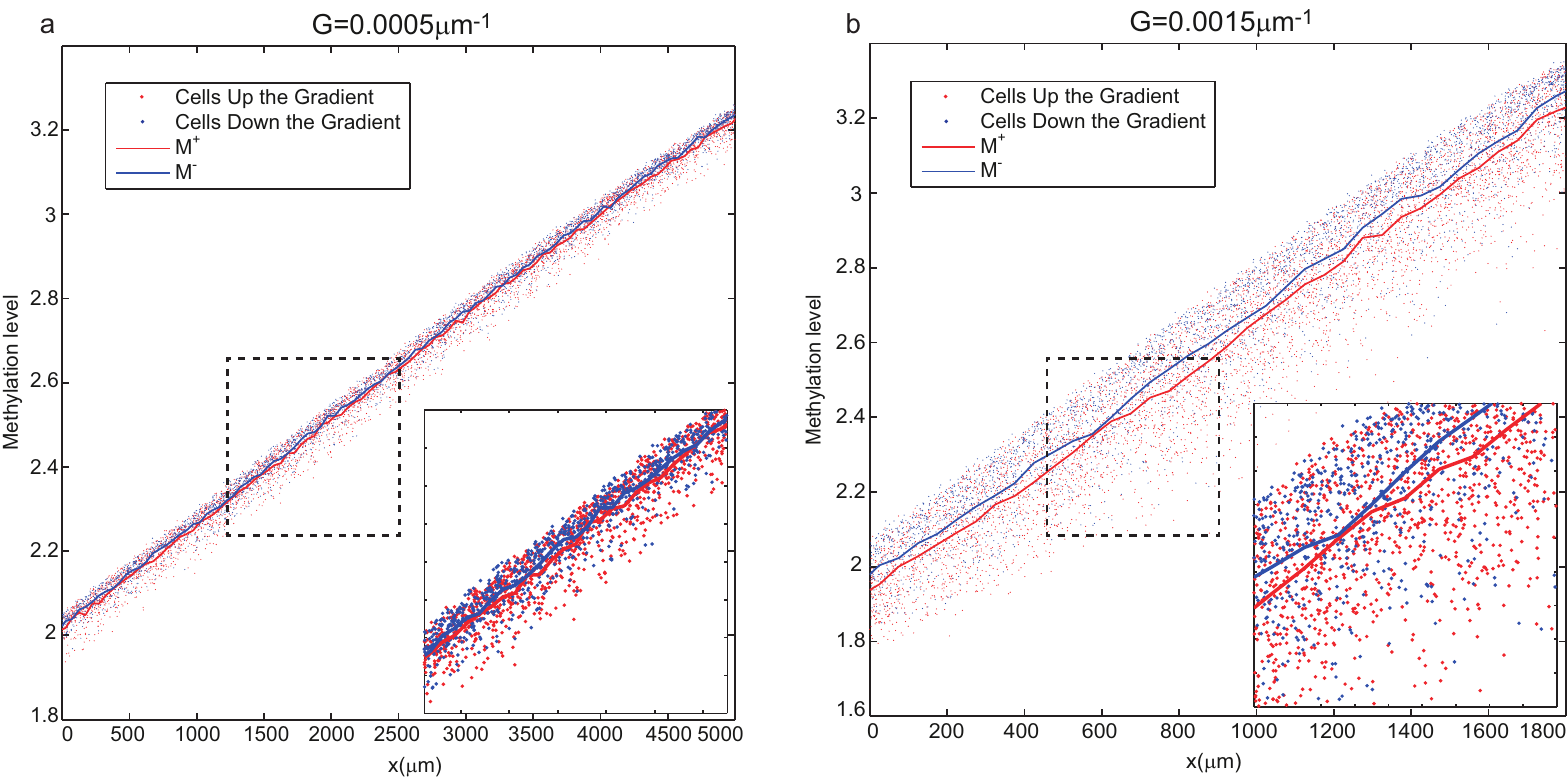}
\caption{The distribution of cells' receptor methylation level for
$G=0.0005\mu m^{-1}$ (a) and $G=0.0015\mu m^{-1}$ (b).
The red dots represent cells moving to right while the blue ones
represent those moving to left. $M^\pm$ are the average methylation
levels for the right and left moving cells
respectively. 
In the simulation, we take $[L]_0 = 5K_I$. Other parameters used in
the SPECS are the same as those proposed in \cite{JiOuTu:10}.
}\label{fig:distribution}
\end{figure}

To justify the Assumption~\ref{asp:closure}, we simulate the
distribution of $m$ with SPECS in an exponential gradient ligand
environment $[L]=[L]_0\exp(Gx)$. SPECS is a well developed
agent-based \emph{E. coli} simulator that incorporates the
physically measured signaling pathways and parameters. We refer the
readers to \cite{JiOuTu:10} for its detailed description. In the
simulation, cells exiting at one side of the boundary will enter
from the other side, and the methylation level is reset randomly
following the local distribution of $m$ at the boundaries. Under
this boundary condition, the system will reach the steady state
after a period of transient process. The steady state distributions
are shown in Figure \ref{fig:distribution}. In each of the
subfigures, the horizontal and vertical axes represent the position
and the methylation level respectively. As shown in Figure
\ref{fig:distribution}, the distribution of methylation level is
localized, and becomes wider when $G$ increases. $M^\pm=\int m P^\pm
\ud m$ are the average methylation levels for the right and left
moving cells. One can also observe that $M^+<M^-$ in the exponential
increasing ligand concentration environment. This can be understood
intuitively by noticing that the up gradient cells with lower
methylation level come from left while the down gradient cells with
higher methylation level come from right.

\begin{figure}[h]
\centering
\includegraphics[width=5.0in]{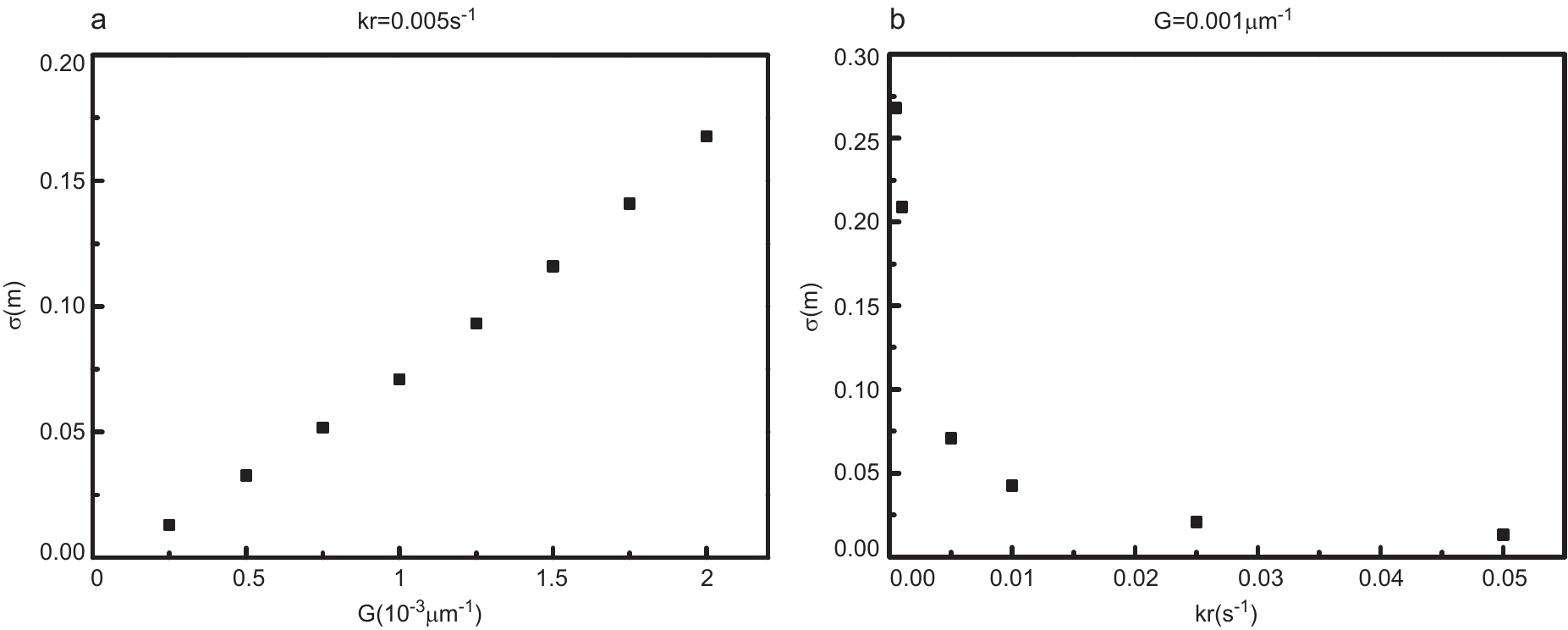}
\caption{The maximum variances $\sigma$ of $m$ for different $G$
and $k_R$. $\sigma$ increases in $G$ for a given
$k_R$ (a) and decreases in $k_R$ with fixed $G$ (b),
but they are all small in the parameter regime we are interested
in.}\label{fig:Gkr}
\end{figure}

In the exponential environment, the numerical variations of $m$ are
almost uniform in $x$.
The maximum of the methylation level variation in the simulation
domain is defined by
$$\sigma \equiv \max{\sqrt{\f{\int\bigl(m-M(x)\bigr)^2(P^+ + P^-) dm}{\int
(P^+ + P^-) dm}}}. $$ Assumption~\ref{asp:closure} is equivalent
to the condition $\sigma \ll 1$.
As shown in Figure \ref{fig:Gkr}, $\sigma$ increases in $G$ and
decreases in $k_R$, but it is always small
in the parameter regime we are interested in, \ie
Assumption~\ref{asp:closure} holds in these cases.

\subsection{The localization of $P^\pm$ in $m$ in the limit of $k_R \gg 1$.}
We show by formal analysis that the assumption $\int (m-M)^2P^\pm
\ud m \ll 1$ is true when the adaptation rate $k_R\gg 1$. Denote
\begin{equation}\label{eq:feta}
k_R=1/\eta, \qquad f(a)=f_\eta(a)/\eta ,
\end{equation}
then \eqref{eq:kineticmodel1}-\eqref{eq:kineticmodel2} become
\begin{eqnarray}
&&\f{\p P^+}{\p t}=-\f{\p(v_0 P^+)}{\p x}-\f{1}{\eta}\f{\p
(f_\eta(a)P^+)}{\p m}-\f{z}{2}(P^+-P^-),
\label{eq:kineticmodel1eta}\\
&&\f{\p P^-}{\p t}=\f{\p(v_0 P^-)}{\p x}-\f{1}{\eta}\f{\p
(f_\eta(a)P^-)}{\p m}+\f{z}{2}
(P^+-P^-).\label{eq:kineticmodel2eta}
\end{eqnarray}
Integrating the above two equations with respect to $m$ produces,
for $P^{\pm}_R(t,x)=\int_0^{R}P^{\pm}(t,x,m)\,dm$ ($R$ is an
arbitrary positive constant),
\begin{align}
&\f{\p P^+_R}{\p t}=-\f{\p(v_0 P^+_R)}{\p
x}-\f{1}{2}\int_0^{R}z(P^+ - P^-) \ud m \label{eq:kineticmodel1pm}
\\&\hspace{5em}
 -\f{1}{\eta}f_\eta\bigl(a(R)\bigr)P^{+}(t,x,R)+
\f{1}{\eta}f_\eta\bigl(a(0)\bigr)P^{+}(t,x,0), \nonumber
\\
&\f{\p P^-_R}{\p t}=\f{\p(v_0 P^-_R)}{\p
x}+\f{1}{2}\int_0^{R}z(P^+ - P^-) \ud m \label{eq:kineticmodel2pm}
\\&\hspace{5em}
-\f{1}{\eta}f_\eta\bigl(a(R)\bigr)P^{-}(t,x,R)+
\f{1}{\eta}f_\eta\bigl(a(0)\bigr)P^{-}(t,x,0) .\nonumber
\end{align}
The probability density functions satisfy $P^{\pm}(t,x,m)\geq 0$,
$\forall m\geq 0$, and thus $P^{\pm}_R(t,x)$ increases with $R$.

We consider the regime
\begin{equation}
\eta\ll1, \quad \text{and} \quad f_\eta(a)\sim \Or(1).
\end{equation}
Then when $\eta\ll 1$,
\eqref{eq:kineticmodel1pm}-\eqref{eq:kineticmodel2pm} indicate for
$R\in(0,+\infty)$, \beq\label{eq:Feta}
f_\eta\bigl(a(R)\bigr)P^{\pm}(t,x,R)=f_\eta\bigl(a(0)\bigr)P^{\pm}(t,x,0)+O(\eta).
\eeq We show by contradiction that when $\eta\to 0$, the boundary
condition at $m=0$ has to satisfy
$f_\eta\bigl(a(0)\bigr)P^{\pm}(t,x,0)\to 0$, $\forall (x,t)\in
\mathbb{R}\times(0,+\infty)$. Otherwise, assume that
\begin{equation}\label{eq:asmp}
f_\eta\bigl(a(0)\bigr)P^{\pm}(t,x,0)\rightarrow
 C(t,x)\neq 0,\quad
 \mbox{for some } (x,t)\in \mathbb{R}\times(0,+\infty).
\end{equation}
Define \beq\label{eq:ma0}
M_{a_0}=\f{1}{\alpha}\Bigl(-\f{1}{N}\ln\bigl(\f{1}{a_0}-1\bigr)
+\ln\bigl(\f{1+[L]/K_I}{1+[L]/K_A}\bigr)\Bigr)+m_0. \eeq Then
\eqref{eq:Fa}-\eqref{eq:am} imply
\begin{align*}
& f_\eta\bigl(a(R)\bigr)>0, \quad \text{when} \quad 0<R<M_{a_0}; \\
& f_\eta\bigl(a(R)\bigr)<0, \quad \text{when} \quad R>M_{a_0}.
\end{align*}
Besides that, one has $P^\pm(t,x,R) >0$, thus
$f_\eta(a(R))P^{\pm}(t,x,R)$ will change sign for different $R$. On
the other hand, when $\eta \ll 1$, \eqref{eq:Feta}, \eqref{eq:asmp}
imply that for $\forall R\in(0,+\infty)$,
$f_\eta\bigl(a(R)\bigr)P^{\pm}(t,x,R)$ has the same sign as
${f_\eta\bigl(a(0)\bigr)}P^{\pm}(t,x,0)$, which is a contradiction.
Therefore, $f_\eta(a(0))P^{\pm}(t,x,0)\to 0$, and as
$\eta\rightarrow 0$, \beq\label{eq:Fetaorder}
f_\eta(a(R))P^{\pm}(t,x,R)\rightarrow 0,\qquad \forall
R\in(0,+\infty) .\eeq Then the definition of $f(a)$ in
\eqref{eq:Fa}-\eqref{eq:am} gives that if $R\neq M_0$,
$P^{\pm}(t,x,R)\to 0$, which implies when $\eta\to 0$,
\begin{equation}\label{eq:dltasp}
P^{\pm}(x,t,m)=P^{\pm}_m\delta(m-M_0).
\end{equation}

\section{Keller-Segel limit and connections to the original PBMFT}\label{sec:KS}
In this section, we derive the Keller-Segel limit from
\eqref{eq:density}-\eqref{eq:momentumflux} by taking into account
the different physical time scales of the tumbling, adaptation and
experimental observations. We shall also connect the new moment
system to the original PBMFT developed in \cite{SiWuOuTu:12}.
Moreover, a numerical comparison of the moment system
\eqref{eq:density}-\eqref{eq:momentumflux} with SPECS is provided in
the environment of spatial-temporally varying concentration.

\subsection{Keller-Segal limit by the parabolic scaling}
We nondimensionalize the moment system
\eqref{eq:density}-\eqref{eq:momentumflux} by letting
$$
t=T\tilde{t},\qquad x=L\tilde{x},\qquad v_0=s_0\tilde{v}_0,
$$
where $T$, $L$ are temporal and spatial scales of the system
respectively. Then
$$J_\rho=s_0\tilde{J}_\rho,\qquad J_q=s_0\tilde{J}_q,$$
and the system becomes (after dropping the ``$\sim$'' )
\begin{align}
\f{1}{T}\f{\p \rho}{\p t}&=-\f{s_0}{L}\f{\p J_\rho}{\p
x},\nonumber
\\
\f{s_0}{T}\f{\p J_\rho}{\p t}&=-v_0^2\f{\p\rho}{\p
x}\f{s_0^2}{L}-\f{s_0}{T_1}ZJ_\rho -\f{s_0}{T_1}\f{\p Z}{\p
m}(J_q-MJ_\rho),\nonumber
\\
\f{1}{T}\f{\p q}{\p t}&=-\f{s_0}{L}\f{\p J_q}{\p
x}+\f{1}{T_2}F\rho,\nonumber
\\
\f{s_0}{T}\f{\p J_q}{\p t}&=-v_0^2\f{\p q}{\p
x}\f{s_0^2}{L}+FJ_\rho\f{s_0}{T_2} +\f{\p F}{\p
m}(J_q-MJ_\rho)\f{s_0}{T_2}-ZJ_q\f{s_0}{T_1}-\f{\p Z}{\p
m}M(J_q-MJ_\rho)\f{s_0}{T_1}, \nonumber
\end{align}
where $T_1$, $T_2$ are the average run and adaptation time scales
respectively.

For \emph{E. coli}, the average run time is at the order of $1$s,
the adaptation time is approximately $10s \sim 100$s, and
according to the experiment in \cite{Zhuetal:12}, the system time
scale when Keller-Segel equation is valid is about $1000$s.
Therefore, we consider the long time regime, where the tumbling
frequency becomes large (the so-called parabolic scaling).
Let
\begin{equation}\label{eq:veps}
\frac{T_1}{L/s_0}=\veps, \quad \frac{T_2}{L/s_0}=1, \quad \text{and}
\quad \frac{T}{L/s_0}=\f{1}{\veps}
\end{equation}
Then \eqref{eq:density}-\eqref{eq:momentumflux} become
\begin{align}
&\veps\f{\p \rho}{\p t}=-\f{\p J_\rho}{\p x},\label{eq:densityscale}\\
 &\veps\f{\p J_\rho}{\p t}=-v_0^2\f{\p \rho}{\p x}-\f{Z}{\veps}J_\rho
 -\f{1}{\veps}\f{\p Z}{\p m}(J_q-MJ_\rho),\label{eq:densityfluxscale}\\
 &\veps\f{\p q}{\p t}=-\f{\p J_q}{\p x}+F\rho,\label{eq:momentumscale}\\
 &\veps\f{\p J_q}{\p t}=-v_0^2\f{\p q}{\p x}+FJ_q+\f{\p F}{\p m}(J_q-MJ_\rho)
 -\f{Z}{\veps}J_q-\f{1}{\veps}\f{\p Z}{\p m}M(J_q-MJ_\rho)\label{eq:momentumfluxscale}.
 \end{align}
Consider the following asymptotic expansion
\begin{align*}
&\rho=\rho^{(0)}+\veps\rho^{(1)}+\cdots, &
J_\rho=J_\rho^{(0)}+\veps J_\rho^{(1)}+\cdots;\\
&q=q^{(0)}+\veps q^{(1)}+\cdots, & J_q=J_q^{(0)}+\veps
J_q^{(1)}+\cdots; \\
&M=M_0+\veps M_1+\cdots, & F=F_0+\veps F_1+\cdots, \\
&Z=Z_0+\veps Z_1+\cdots.
\end{align*}
Matching the $O(1/\veps)$ terms in \eqref{eq:densityfluxscale} and
\eqref{eq:momentumfluxscale} gives
\begin{equation*}
Z_0 J_\rho^{(0)}=\f{\p Z_0}{\p
m}(M_0J_\rho^{(0)}-J_q^{(0)}), \quad \text{and} \quad
Z_0 J_q^{(0)}=\f{\p Z_0}{\p
m}M_0(M_0J^{(0)}_\rho -J^{(0)}_q),
\end{equation*}
which implies
\begin{equation*}
M_0J_\rho^{(0)}=J_q^{(0)}, \quad \text{and} \quad
J_\rho^{(0)}=J_q^{(0)}=0.
\end{equation*}
Hence the $O(1)$ term in \eqref{eq:momentumscale} yields
\begin{equation*}
F_0=0.
\end{equation*}
Equating the $O(\veps)$ terms in \eqref{eq:densityscale} and
\eqref{eq:momentumscale} produces \beq\label{eq:order0rho} \f{\p
\rho^{(0)}}{\p t}=-\f{\p J_\rho^{(1)}}{\p x}, \quad \text{and}
\quad \f{\p q^{(0)}}{\p t}=-\f{\p J^{(1)}_q}{\p
x}+F_1\rho^{(0)}. \eeq Putting together the $O(1)$ terms in
\eqref{eq:densityfluxscale} and \eqref{eq:momentumfluxscale}
brings
\begin{align}
\label{eq:order1densityflux} & -v_0^{2}\f{\p \rho^{(0)}}{\p
x}-Z_0 J_\rho^{(1)}+\f{\p Z_0}{\p
m}(M_0J_\rho^{(1)}-J_q^{(1)})=0 \\
\label{eq:order1momentumflux} & -v_0^2\f{\p q^{(0)}}{\p x}-Z_0
J_q^{(1)}-\f{\p Z_0}{\p
m}M_0(J_q^{(1)}-M_0J_\rho^{(1)})=0.
\end{align}
The above two equations imply
\begin{equation}
\begin{aligned}
J_\rho^{(1)}&=Z_0^{-1}\biggl(-v_0^{2}\f{\p \rho^{(0)}}{\p
x}+\f{\p Z_0}{\p m}
\Bigl(M_0J_\rho^{(1)}-J_q^{(1)}\Bigr)\biggr)
\\
&=-Z_0^{-1}v_0^2\f{\p\rho^{(0)}}{\p x}+Z_0^{-2}\f{\p
Z_0}{\p m}
\biggl(-v_0^{2}\Bigl(M_0\f{\p \rho^{(0)}}{\p x}-\f{\p q^{(0)}}{\p x}\Bigr)\biggr)\\
&=-Z_0^{-1}v_0^2\biggl(1+(Z_0)^{-1}M_0\f{\p Z_0}{\p
m}\biggr)\f{\p\rho^{(0)}}{\p x} +Z_0^{-2}v_0^2\f{\p
Z_0}{\p m}\f{\p q^{(0)}}{\p x}
\end{aligned}
\end{equation}
By \eqref{eq:M},
\begin{equation*}
q^{(0)}=M_0\rho^{(0)}.
\end{equation*}
Therefore
\begin{equation}\label{eq:Jrho1}
\begin{aligned}
J_\rho^{(1)}&=-Z_0^{-1}v_0^2\f{\p\rho^{(0)}}{\p
x}+Z_0^{-2}v_0^{2}\f{\p Z_0}{\p m} \big(\f{\p q^{(0)}}{\p
x}-M_0\f{\p \rho^{0}}{\p x}\big)\\&=
-Z_0^{-1}v_0^2\f{\p\rho^{(0)}}{\p
x}+Z_0^{-2}v_0^{2}\rho^{(0)}\f{\p Z_0}{\p m}\f{\p
M_0}{\p x}.
\end{aligned}
\end{equation}
Substituting \eqref{eq:Jrho1} into \eqref{eq:order0rho} gives the
K-S equation
\begin{equation}\label{eq:KSlimit}
\f{\p \rho^{(0)}}{\p t}=v_0^2\f{\p }{\p
x}\biggl(Z_0^{-1}\f{\p\rho^{(0)}}{\p x}\biggr) -v_0^2\f{\p}{\p
x}\biggl(Z_0^{-2}\f{\p Z_0}{\p m}\f{\p M_0}{\p
x}\rho^{(0)}\biggr).
\end{equation}
Using \eqref{eq:Z}, \eqref{eq:ma0} and $M_0=M_{a_0}$,
$Z_0=z(M_{a_0})$, the K-S equation becomes \beq\label{eq:KSrho}
\f{\p \rho^{(0)}}{\p t}=v_0^2\f{\p }{\p
x}\big(Z_0^{-1}\f{\p\rho^{(0)}}{\p x}\big) -\f{\p}{\p
x}\big(\chi_0\rho^{(0)}\f{\p f_0}{\p x}\big) \eeq with
$\displaystyle \chi_0=\f{v_0^2
\tau^{-1}}{(z_0+\tau^{-1})^2}NH(1-a_0)$.

\begin{remark}
1. Instead of \eqref{eq:veps}, if we consider
$$
\frac{T_1}{L/s_0}=\veps, \quad \frac{T_2}{{L/s_0}}=\kappa\veps,
\quad \text{and} \quad \frac{T}{{L/s_0}}=\f{1}{\veps},
$$
then the rescaled system becomes
\begin{align}
\nonumber\veps\frac{\partial \rho}{\partial t}& =-\frac{\partial
J_\rho}{\partial x}, \\
\nonumber \veps\f{\p J_\rho}{\p t}&=-v_0^{2}\f{\p \rho}{\p
x}-\f{Z}{\veps}J_\rho-\f{1}{\veps}\f{\p Z}{\p m}(J_q-MJ_\rho), \\
\nonumber \veps\f{\p q}{\p t}&=-\f{\p J_q}{\p x}+\f{1}{\kappa\veps}F\rho, \\
\nonumber\veps\f{\p J_q}{\p t} &= -v_0^{2}\f{\p q}{\p
x}+\f{1}{\kappa\veps}F J_\rho+\f{1}{\kappa\veps}\f{\p F}{\p
m}(J_q-MJ_\rho)-\f{Z}{\veps}J_q -\f{1}{\veps}\f{\p Z}{\p
m}M(J_q-MJ_\rho).
\end{align}
When $\kappa\leq O(1/\veps)$, carrying on similar asymptotic
expansion will produce the same Keller-Segel limit \eqref{eq:KSrho}
as $\epsilon\to 0$. This indicates that when the adaptation time is
shorter than $\sqrt{TT_1}$, the Keller-Segel equation is valid for
\emph{E. coli} chemotaxis.

2. The velocity scale of individual bacteria is $s_0$. The temporal
and spacial scales of the system we consider are $T$ and $L$
respectively, therefore the velocity scale of the drift velocity
$v_d={J_\rho}/{\rho}$ is $L/T$. The equation \eqref{eq:veps} implies
$v_d/s_0\sim O(\veps)$, which means that in the regime where K-S
equation is valid, the drift velocity is much smaller than the
moving velocity of individual bacteria.
\end{remark}

\subsection{Connection to the original PBMFT}
We shall show that, under certain assumptions, the moment system
\eqref{eq:density}-\eqref{eq:momentumflux} gets to the
original PBMFT in \cite{SiWuOuTu:12}. 
Especially, one of the equations delivers the important physical
assumption eqn. $(3)$ in \cite{SiWuOuTu:12}.

The macroscopic quantities in the PBMFT in \cite{SiWuOuTu:12} are
the cell densities to the right $P^+$ and to the left $P^-$, and the
total density $\rho_s=P^++P^-$ and cell flux $J_s=v_0(P^+-P^-)$; the
average methylation level to the right $M^+$ and to the left $M^-$,
the methylation difference $\Delta M_s=\f{1}{2}(M^+-M^-)$ and the
average methylation $M_s=\f{M^+P^++M^-P^-}{P^++P^-}$. The model in
\cite{SiWuOuTu:12} is
\begin{align}\label{eqn:rho}
& \frac{\partial \rho_s}{\partial t}=-\frac{\partial J_s}{\partial
x},
\\ \label{eqn:J}
& \frac{\partial J_s}{\partial t}\approx -v_0^2\frac{\partial
\rho_s}{\partial x}-ZJ_s-v_0\frac{\partial Z}{\partial m}\Delta
M_s\rho_s,
\\ \label{eqn:M}
& \frac{\partial M_s}{\partial t}\approx
F-\frac{J_s}{\rho_s}\frac{\partial M_s}{\partial
x}-\frac{1}{\rho_s}\frac{\partial}{\partial x}(v_0\Delta
M_s\rho_s),
\end{align}
together with the physical assumption 
\beq\label{eqn:deltaM} \Delta M_s \approx -\f{\p M_s}{\p
x}Z^{-1}v_0, \eeq  which physically means $\Delta M_s$ is
approximated by the methylation level difference in the mean
methylation field $M_s(x, t)$ over the average run length
$v_0Z^{-1}$, due to the fact that the direction of motion is
randomized during each tumble event.

We firstly discuss the connections of
\eqref{eq:rhojrho}-\eqref{eq:qjq} to the macroscopic quantities in
\eqref{eqn:rho}--\eqref{eqn:deltaM}. By definition, one has
\begin{align}
& \rho=P^++P^-=\rho_s, \quad J_\rho=v_0(P^+-P^-)= J_s, \\ & q=M^+P^++M^-P^-=M_s\rho_s, \\
& J_q=v_0\big(M^+P^+-M^-P^-\big)=M_sJ_s+\frac{v_0\Delta
M_s}{\rho_s}\biggl(\rho_s^2-\frac{J_s^2}{v_0^2}\biggr).\label{eq:Jqs}
\end{align}

\begin{assump}\label{asp:vdv0}
Defining the drift velocity $v_d=J_s/\rho_s$, we assume
\begin{equation*}
\abs{v_d}\ll v_0.
\end{equation*}
\end{assump}
This is a fair assumption since in experiments an individual cell
usually travels at a much higher speed than that of chemotaxis.

Applying Assumption~\ref{asp:vdv0} in \eqref{eq:Jqs} gives
\begin{equation}\label{eq:MJrhoJq}
J_q - MJ_\rho\approx v_0\Delta M_s\rho_s,
\end{equation}
by which, \eqref{eq:density}, \eqref{eq:densityflux} and
\eqref{eq:momentum} become \eqref{eqn:rho}, \eqref{eqn:J} and
\eqref{eqn:M} respectively.
In particular, substituting \eqref{eq:MJrhoJq} into
\eqref{eq:momentumflux} produces
\begin{multline}
\frac{\partial}{\partial t}(M_sJ_s+v_0\Delta
M_s\rho_s)\approx -v_0^2\frac{\partial}{\partial x}(\rho_s M_s)\\
+\biggl(FJ_s+v_0\frac{\partial F}{\partial m}\Delta M_s
\rho_s\biggr)-\biggl(M_sZJ_s+v_0\frac{\partial (M_sZ)}{\partial
m}\Delta M_s\rho_s\biggr).
\end{multline}
Using \eqref{eqn:rho}-\eqref{eqn:M} yields
\begin{multline}\label{eq:deltaM}
v_0\rho_s\frac{\partial \Delta M_s}{\partial
t}\approx\frac{J^2_s}{\rho_s}\frac{\partial M_s}{\partial
x}+\frac{J_s}{\rho_s}\frac{\partial}{\partial x}(v_0\rho\Delta M_s)
\\ +v_0\Delta M_s\frac{\partial J_s}{\partial x}-v_0^2\rho_s\frac{\partial
M_s}{\partial x}+v_0\rho_s\Delta M_s\biggl(\frac{\partial
F}{\partial M}-Z\biggr).
\end{multline}

Assumption~\ref{asp:vdv0} formally implies the terms containing
$J_s$ are relatively small in \eqref{eq:deltaM}, and by a
quasi-static approximation $\partial \Delta M_s/\partial t\approx
0$, one has the sum of last three terms is approximately zero in
\eqref{eq:deltaM}, and thus
\begin{equation*}
\Delta M_s \approx v_0\frac{\partial M_s}{\partial
x}\frac{1}{\frac{1}{\rho_s}\frac{\partial J_s}{\partial
x}+\frac{\partial F}{\partial m}-Z}.
\end{equation*}
When $ Z>>\abs{\frac{1}{\rho_s}\frac{\partial J_s}{\partial
x}+\frac{\partial F}{\partial m}}$, the above equation leads to the
important physical assumption \eqref{eqn:deltaM},
\begin{equation}\label{eq:dMinSi}
\Delta M\approx -\f{\p M}{\p x}Z^{-1}v_0,
\end{equation}
which recovers the PBMFT model in \cite{SiWuOuTu:12}.
\begin{remark}
The model \eqref{eqn:J}--\eqref{eqn:deltaM} is a nonlinear
advection-diffusion system. The new moment system
\eqref{eq:density}-\eqref{eq:momentumflux} evolves only linear
advection terms and nonlinear reactions, and thus the numerical
methods for such a system is well studied \cite{Le:92}.
\end{remark}

\subsection{Numerical comparison to SPECS}
To show the validity of the moment system
\eqref{eq:density}-\eqref{eq:momentumflux}, numerical comparisons to
SPECS will be presented in this subsection. We choose
spatial-temporal varying environment to show how the intracellular
dynamics affects the \emph{E. coli} behaviors at the population
level. Specifically it presents a pattern of traveling attractant
concentration wave, in which an interesting reversal of chemotaxis
group velocity was revealed in \cite{SiWuOuTu:12}.

\begin{figure}
\centering
\includegraphics[width=5.0in]{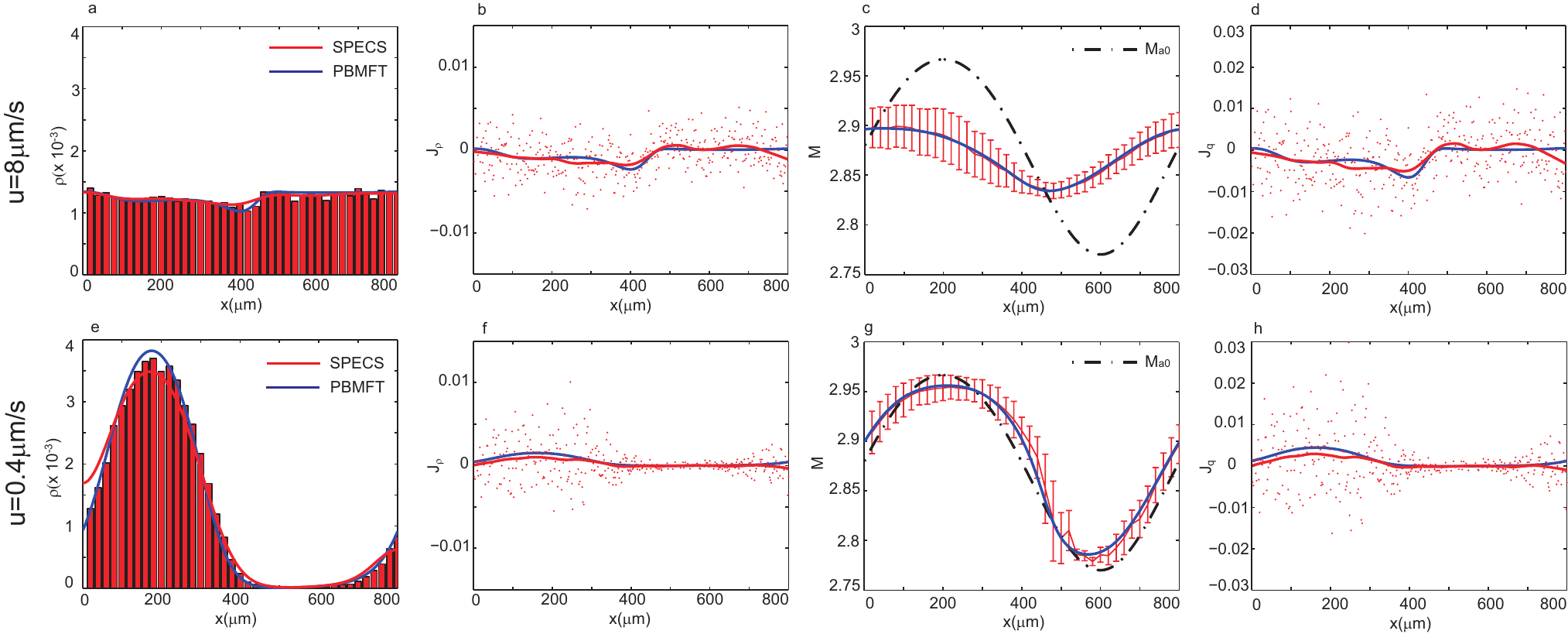}
\caption{Numerical comparison between the new moment system of PBMFT
and SPECS. The steady state profiles of $\rho$: (a, e), $J_\rho$:
(b, f), $M=q/\rho$: (c, g), $J_q$: (d, h) when the traveling wave
speeds are  $u=8\mu m/s$ and $u=0.4\mu m/s$ respectively. In the
subfigures, red lines and dots are from SPECS (red lines in b, d, f,
h are the smoothed results of the red dots), while blue lines are
from the new moment system of PBMFT. Parameters used here are $[L]_0
= 500\mu M$, $[L]_A = 100\mu M$, $\lambda = 800 \mu m$. $20000$
cells are simulated in SPECS. }\label{fig:numerical}
\end{figure}

We consider a circular channel with the travelling wave
concentration given by $[L](x,t) = [L]_0 +[L]_A +
\sin[\f{2\pi}{\lambda}(x-ut)]$. The wavelength $\lambda$ is fixed to
be the length of the channel, while the wave velocity $u$ can be
tuned. The steady state profiles of all the macroscopic quantities
in \eqref{eq:density}-\eqref{eq:momentumflux} and corresponding
SPECS results are compared in Figure \ref{fig:numerical}. The
results from SPECS and the moment system are quantitatively
consistent. It can be noticed that, when the concentration changes
slowly ($u = 0.4\mu m/s$), the profile of $M$ can catch up with the
target value $M_{a_0}$ (defined by $a([L], M_{a_0}) = a_0$), while
in the fast-varying environment ($u=8\mu m/s$) there is a lag in
phase between $M$ and $M_{a_0}$. This difference is caused by the
slow adaptation rate of cell and it also leads to the difference in
the profiles of $\rho$ and even chemotaxis velocity; we refer
interested readers to \cite{SiWuOuTu:12} for more detailed
discussions and physical explanations.

\section{Two dimensional mean-field model}\label{sec:2dMFT}
In this section, we derive the two-dimensional moment system of PBMFT based on a
formal argument using the point-mass assumption in methylation and
the minimization principle proposed in \cite{Hi:02}.

In two dimensions, $\bd{v}=v_0(\cos\theta,\sin\theta)$, where $v_0$ is
the velocity magnitude. $P(t,\bd{x},\bd{v},m)$ in
\eqref{eq:kinetic2d} can be rewritten as $P(t,\bd{x},\theta,m)$.
$z(m,[L],\theta,\theta')$ is the tumbling rate from $\theta'$ to
$\theta$. The tumbling term $Q(P,z)$ in \eqref{eq:Q} becomes
\begin{equation}\label{eq:Q2}
Q(P,z)=\barint_V
z(m,[L],\theta,\theta')P(t,\bd{x},\theta',m)\ud\theta' -\barint_V
z(m,[L],\theta',\theta)\ud\theta'P(t,\bd{x},\theta,m),
\end{equation}
where $V=[0,2\pi)$ and $\barint=\frac{1}{2\pi}\int_V$. According
to \eqref{eq:Z}, $z(m,[L],\theta,\theta')$ is independent of
$\theta$ and thus we denote it by $z(m,[L])$.



Define \beq\label{eq:fg} g(t,\bd{x},\theta)=\int
P(t,\bd{x},\theta,m)\ud m, \qquad h(t,\bd{x},\theta)=\int
mP(t,\bd{x},\theta,m)\ud m; \eeq \beq\label{eq:Mbar}
M(t,\bd{x},\theta)=\f{h(t,\bd{x},\theta)}{g(t,\bd{x},\theta)},\qquad
\overline{M}(t,\bd{x})=\f{\int_Vh(t,\bd{x},\theta)\ud\theta}{\int_Vg(t,\bd{x},\theta)\ud\theta};
\eeq and the density, density flux, momentum (in $m$), and momentum
flux as follows
\begin{align}
\rho(t,\bd{x})=\barint_Vg(t,\bd{x},\theta)\ud\theta,&
 \quad J_\rho(t,\bd{x})=\barint_V\bd{v}g(t,\bd{x},\theta)\ud\theta;
 \label{eq:rhojrho2}\\
q(t,\bd{x})=\barint_Vh(t,\bd{x},\theta)\ud\theta,&
 \quad
 J_q(t,\bd{x})=\barint_V\bd{v}h(t,\bd{x},\theta)\ud\theta.
 \label{eq:qjq2}
\end{align}
We assume \beq\label{eq:P2d}
P(t,\bd{x},\theta,m)=g(t,\bd{x},\theta)\delta\bigl(m-M(t,\bd{x},\theta)\bigr).
\eeq This assumption is motivated by \eqref{eq:dltasp} in one
dimension, which could be formally understood as the limit of
$k_R\rightarrow +\infty$.

Denote
$$Z=z(\overline{M},[L]),\quad
\frac{\p Z}{\p M}=\f{\p z}{\p m}(\overline{M},[L]),\quad
F=f(\overline{M},[L]),\quad \frac{\p F}{\p M}=\f{\p f}{\p
m}(\overline{M},[L]).
$$

Integrating \eqref{eq:kinetic2d} with respect to $m$ yields
\begin{equation}\label{eq:fm}
{\p_t g}=-\bd{v}\cdot\nabla_{\bd{x}} g
+\barint_Vz\bigl(M(\theta'),[L]\bigr)g(t,\bd{x},\theta')\ud\theta'
-z\bigl(M(\theta),[L]\bigr)g(t,\bd{x},\theta).
\end{equation}
Integrating \eqref{eq:fm} with respect to $\theta$ gives the
equation for density, \beq \label{eq:rho2d}
\f{\p\rho(t,\bd{x})}{\p t}=-\nabla_{\bd{x}}\cdot J_\rho. \eeq
Multiplying \eqref{eq:fm} by $\bd{v}$ and integrating with respect
to $\theta$ produce
\begin{equation}
\begin{aligned}
\f{\p J_\rho}{\p t}&=-\barint_V\bd{v}\otimes\bd{v}\nabla_{\bd{x}}
g\ud\theta
-\barint_V\bd{v}z(M(\theta),[L])g(x,t,\theta)\ud\theta \\
&\approx-\barint_V\bd{v}\otimes\bd{v}\nabla_{\bd{x}} g\ud\theta-
\barint_V\bd{v}\big(z(\overline{M},[L])+\f{\p Z}{\p
m}(M(\theta)-\overline{M})\big)g(x,t,\theta)\ud\theta
\\
&=-\barint_V\bd{v}\otimes\bd{v}\nabla_{\bd{x}} g\ud\theta
-ZJ_\rho-\f{\p Z}{\p m}\big(J_q-\overline{M}J_\rho\big),
\end{aligned}
\label{eq:Jrho2d}
\end{equation}
where we have used the first order Taylor expansion in the second
step.

Integrating $m\times$\eqref{eq:kinetic2d} with respect to $m$
brings
\begin{multline}\label{eq:gm}
{\p_t h}=-\bd{v}\cdot\nabla_{\bd{x}} h+
f\bigl(M(\theta),[L]\bigr)g(\theta)
+\barint_Vz\bigl(M(\theta'),[L]\bigr)g(\theta')M(\theta')\ud\theta'
\\-z(M(\theta),[L])g(\theta)M(\theta).
\end{multline}

Integrating \eqref{eq:gm} with respect to $\theta$, and using the
definition in \eqref{eq:Mbar} give
\begin{equation}
\begin{aligned}
\f{\p q(\bd{x},t)}{\p t}&=-\nabla_{\bd{x}}\cdot
J_q+\barint_Vf(M(\theta),[L])g(\bd{x},t,\theta)\ud\theta
\\
&\approx-\nabla_{\bd{x}}\cdot
J_q+\barint_V\big(f(\overline{M},[L])+\f{\p f}{\p
m}(M(\theta)-\overline{M})\big)
g(\bd{x},t,\theta)\ud\theta\\
&=-\nabla_{\bd{x}}\cdot J_q+F\rho.
\end{aligned}
\label{eq:q2d}
\end{equation}
Finally, we integrate $\bd{v}\times$\eqref{eq:gm} with respect to
$\theta$,
\begin{equation}\label{eq:Jq2d}
\begin{aligned} \f{\p J_q}{\p
t}&=-\barint_V\bd{v}\otimes\bd{v}\cdot\nabla_{\bd{x}} h\ud\theta
+\barint_V\bd{v}\bigl(f(M(\theta),[L])-z(M(\theta),[L])M(\theta)\bigr)g(\theta)\ud\theta
\\
&=-\barint_V\bd{v}\otimes\bd{v}\cdot\nabla_{\bd{x}} h\ud\theta
+FJ_\rho+\f{\p F}{\p m}(J_q-\overline{M}J_\rho)-ZJ_q -\f{\p Z}{\p
m}\overline{M}(J_q-\overline{M}J_\rho).
\end{aligned}
\end{equation}

In order to close the system, especially to find equations for
$\rho$, $J$, $q$ and $J_q$, we need to a constitutive relation that
represents $\int_V\bd{v}\otimes\bd{v}\cdot\nabla_{\bd{x}}g\ud\theta$
and $\int_V\bd{v}\otimes\bd{v}\cdot\nabla_{\bd{x}}h\ud\theta$ by
$\rho$, $J$, $q$ and $J_q$. The minimization principle proposed in
\cite{Hi:02} (mathematically a projection of $g$ and $h$ on the
linear space spanned by $1$ and $\bd{v}$) gives \beq\label{eq:gM}
\begin{array}{rl}
g(t,\bd{x},\theta)&\approx g_1(t,\bd{x})+g_c(t,\bd{x})\cos\theta+g_s(t,\bd{x})\sin\theta,\\
h(t,\bd{x},\theta)&\approx
h_1(t,\bd{x})+h_c(t,\bd{x})\cos\theta+h_s(t,\bd{x})\sin\theta.
\end{array}
\eeq Then from \eqref{eq:rhojrho}, \eqref{eq:qjq},
\begin{eqnarray}
\rho(t,\bd{x})&\approx&\barint_V\big(g_1+g_c\cos\theta+g_s\sin\theta\big)\ud\theta=g_1,
\nonumber\\
J_\rho(t,\bd{x})&\approx&\barint_V\bd{v}g(t,\bd{x},\theta)\ud\theta
=\frac{v_0}{2}(g_c, g_s)^\TT,
\nonumber\\
q(t,\bd{x})&\approx&\barint_V\big(h_1+h_c\cos\theta+h_s\sin\theta\big)\ud\theta
=h_1\nonumber\\
J_q(t,\bd{x})&\approx&\barint_V\bd{v}\big(h_1+h_c\cos\theta+h_s\sin\theta\big)\ud\theta
=\f{v_0}{2}(h_c,h_s)^\TT. \nonumber
\end{eqnarray}
Therefore, expressing $g_1$, $g_c$, $g_s$, $M_1$, $M_c$, $M_s$ by
$\rho$, $J_\rho$, $q$, $J_q$, we find
\begin{eqnarray}
&&g_1=\rho,\qquad g_c=\f{2J_{\rho,x}}{v_0},\qquad
g_s=\f{2J_{\rho,y}}{v_0}{},\nonumber\\
&&h_1=q,\qquad h_c=\f{2J_{q,x}}{v_0},\qquad
h_s=\f{2J_{q,y}}{v_0}{},\nonumber
\end{eqnarray}
where we denote $\bd{x}=(x,y)$ and $J_{x}$ and $J_{y}$ are the $x$
and $y$ components of $J$. Hence,
\begin{align}
\barint_V\bd{v}\otimes\bd{v}\cdot\nabla g\ud\theta \approx&
\f{v_0^2}{2}\nabla g_1=\f{v_0^2}{2}\nabla\rho,\label{eq:vvg}
\\
\barint_V\bd{v}\otimes\bd{v}\cdot\nabla h\ud\theta \approx&
\f{v_0^2}{2}\nabla h_1 =\f{v_0^2}{2}\nabla q.\label{eq:vvh}
\end{align}
Furthermore, since \beq\overline{M}=
\f{q}{\rho},\label{eq:barM}\eeq we are able to close the system
\eqref{eq:rho2d}, \eqref{eq:Jrho2d}, \eqref{eq:q2d},
\eqref{eq:Jq2d} using \eqref{eq:gM}.

In summary, \eqref{eq:rho2d}, \eqref{eq:Jrho2d},  \eqref{eq:q2d},
\eqref{eq:Jq2d} together with \eqref{eq:vvg}, \eqref{eq:vvh} and
\eqref{eq:barM} give us a two-dimensional moment system of PBMFT
that is similar to \eqref{eq:density}--\eqref{eq:momentumflux}.

%


\section{Discussion and conclusion}\label{sec:conclusion}

In seek a model at the population level that incorporates
intracellular pathway dynamics, we build a new moment system of
PBMFT in this paper by using moment closure technique in kinetic
theory under the assumption that the methylation level is locally
concentrated. The new system is hyperbolic with linear convection
terms. Under certain assumptions on the drift velocity, the new
system recovers the original model in \cite{SiWuOuTu:12}. Especially
the assumption on the methylation difference made in
\cite{SiWuOuTu:12} can be understood explicitly in this moment
system. We show that when the average run time is much shorter than
that of the population dynamics (parabolic scaling), the
hydrodynamic limit of the moment system can be described by the
Keller-Segal model. We also present numerical evidence to show the
quantitative agreement of the moment system with SPECS
(\cite{JiOuTu:10}).

We remark that the idea of incorporating the underlying signaling
dynamics into the classical population level chemotaxi description
has appeared in the pioneer works of Othmer \etal
\cite{ErOt:04,ErOt:05,XuOt:09}. Here, the internal dynamics follows
the physical model proposed in \cite{SiWuOuTu:12}, which results in
a closure strategy different from \cite{ErOt:04,ErOt:05,XuOt:09}.
The major differences between the kinetic model in
\cite{ErOt:04,ErOt:05,XuOt:09} and the one used here are: the
methylation rate function is nonlinear (in the methylation level) in
\cite{SiWuOuTu:12} while linear in \cite{ErOt:04,ErOt:05,XuOt:09};
the tumbling frequency \eqref{eq:Z} follows the results of recent
physical studies on chemotaxis (\cite{MeTu:03, Keetal:06}).

Another interesting behavior related to the chemo-sensory system of
bacteria is the ``volcano effect'' observed numerically in
\cite{BrLeLi:07}. It may be also important to study this phenomena
in a more physical way and understand the communications between
bacteria using the moment closure technique introduced in this
paper, which will be our future study.

\section*{Appendix}
We give a systematic way of obtaining systems with
higher order moments by further introducing \beq
e(x,t)=\int(m-M)^2(P^++P^-)\ud m,\qquad J_e(x,t)=v_0\int
(m-M)^2(P^+-P^-)\ud m, \eeq and finding a system of six variables
$\rho$, $q$, $e$, $J_\rho$, $J_q$ and $J_e$. The calculations are
almost the same as those of deriving \eqref {eq:density},
\eqref{eq:densityflux}, \eqref{eq:momentum} and
\eqref{eq:momentumflux}.

Define $M^+$, $M^-$, $M$, $P_m^{\pm}$, $Z$, $\f{\p Z}{\p m}$, $F$
and $\f{\p F}{\p m}$ the same as in \eqref{eq:M}, \eqref{eq:PF},
and introduce
$$
\f{\p^2 Z}{\p m^2}=\f{\p^2 z}{\p m^2}\big|_{m=M}, \quad \f{\p^2
F}{\p m^2}=\f{\p^2f}{\p m^2}\big|_{m=M}.
$$
The equation for density $\rho$ is the same as in \eqref{eq:density}. Since
we keep more terms in the Taylor approximation for $z(m)$, the
equation for the density flux $J_\rho$ becomes
\begin{equation*}
\begin{aligned}
\f{\p J_\rho}{\p t}&=-v_0^2\f{\p \rho}{\p x}-v_0\int z(m)(P^+ - P^-) \ud m\\
&\approx -v_0^2\f{\p \rho}{\p x}-v_0\int \biggl(z(M)+\f{\p z}{\p
m}\Big\vert_{m=M}(m-M) +\f{1}{2}\f{\p^2 z}{\p m^2}\Big\vert_{m=M}(m-M)^2\biggr)\\
&\quad\cdot(P^+ - P^-) \ud m \\
&=-v^2_0\f{\p \rho}{\p x}-ZJ_\rho+\f{\p Z}{\p m}MJ_\rho-\f{\p
Z}{\p m}J_q-\f12\f{\p^2 Z}{\p m^2}J_e,
\end{aligned}
\end{equation*}
Similarly, the equation for $q$ is
$$
\f{\p q}{\p t}=-\f{\p J_q}{\p x}+F\rho+\f{1}{2}\f{\p^2 F}{\p
m^2}e.
$$
and for $J_q$ is
$$
\begin{aligned}
\f{\p J_q}{\p t}=&-v_0^2\f{\p q}{\p x}+FJ_\rho+\f{\p F}{\p
m}(J_q-MJ_\rho)-ZJ_q -\f{\p Z}{\p m}M(J_q-MJ_\rho)\\
&-\f{1}{2}\Big(-\f{\p^2 F}{\p m^2}+M\f{\p^2 Z}{\p m^2}+2\f{\p
Z}{\p m}\Big)J_e.
\end{aligned}
$$
We need two additional equations for $e$ and $J_e$. Since
$$\begin{aligned}
e(x,t)&=\int (m-M)^2(P^++P^-)\ud m=
\int m^2(P^++P^-)\ud m-2Mq+M^2\rho\\
&=\int m^2(P^++P^-)\ud m-Mq,
\end{aligned}
$$
$$
J_e(x,t)=v_0\int (m-M)^2(P^+-P^-)\ud m= v_0\int m^2(P^+-P^-)\ud
m-2MJ_q+M^2J_\rho,
$$
we can get the equation for $e$ by multiplying both sides of
\eqref{eq:kineticmodel1}+\eqref{eq:kineticmodel1} by $m^2$ and
integrating with respect to $m$,
$$
\begin{aligned}
\f{\p (e+Mq)}{\p t}&=-\f{\p(J_e+2MJ_q-M^2J_\rho)}{\p x}-\int m^2\f{\p \big(f(a)(P^++P^-)\big)}{\p m}\ud m\\
&=
-\f{\p(J_e+2MJ_q-M^2J_\rho)}{\p x}+2\int mf(a)(P^++P^-)\ud m\\
&\approx -\f{\p(J_e+2MJ_q-M^2J_\rho)}{\p x} +2\int
\biggl(MF+\f{\p\big( mf(a)\big)}{\p
m}\Big\vert_{m=M}(m-M) \\
&\quad+\f{1}{2}\f{\p^2\big( mf(a)\big)}{\p m^2}\Big\vert_{m=M}(m-M)^2\biggr)(P^++P^-)\ud m\\
&=-\f{\p(J_e+2MJ_q-M^2J_\rho)}{\p x} +2MF\rho+\big(M\f{\p^2F}{\p
m^2}+2\f{\p F}{\p m}\big)e.
\end{aligned}
$$
The equation for $J_e$ can be obtained by multiplying both sides of
\eqref{eq:kineticmodel1}-\eqref{eq:kineticmodel1} by $v_0m^2$ and
integrating with respect to $m$,
$$
\begin{aligned}
&\f{\p (J_e+2MJ_q-M^2J_\rho)}{\p t}\\
=&-v_0^2\f{\p\big(e+Mq\big) }{\p x} -v_0\int m^2\f{\p
\big(f(a)(P^+-P^-)\big))}{\p m}\ud m-v_0\int z(m)m^2(P^+-P^-)\ud m
\\
=&-v_0^2\f{\p\big(e+Mq\big) }{\p x}
+2v_0\int mf(a)(P^+-P^-)\ud m-v_0\int z(m)m^2(P^+-P^-)\ud m\\
\approx&-v_0^2\f{\p\big(e+Mq\big) }{\p
x}+2v_0\int\biggl((mf(a))\vert_{m=M}+\f{\p
(mf)}{\p m}\Big\vert_{m=M}(m-M) \\
&\quad+\f12\f{\p^2 (mf)}{\p m^2}\Big\vert_{m=M}(m-M)^2
\biggr)(P^+-P^-)\ud m
-v_0\int\biggl((z(m)m^2)\vert_{m=M}\\&\quad+\f{\p (z(m)m^2)}{\p
m}\Big\vert_{m=M}(m-M) +\f12\f{\p^2 (z(m)m^2)}{\p
m^2}\Big\vert_{m=M}(m-M)^2\biggr)(P^+-P^-)\ud m
\\
=&-v_0^2\f{\p\big(e+Mq\big) }{\p x}+2MFJ_\rho+2\big(M\f{\p F}{\p
m}+F\big)\big(J_q-MJ_\rho\big)
+\big(M\f{\p^2F}{\p m^2}+2\f{\p F}{\p m}\big)J_e\\
&\quad-M^2ZJ_\rho-\big(2MZ+M^2\f{\p Z}{\p
m}\big)\big(J_q-MJ_\rho\big) -\f{1}{2}\big(\f{\p^2 Z}{\p
m^2}M^2+4M\f{\p Z}{\p m}+2Z\big)J_e.
\end{aligned}
$$
All these six equations for $\rho$, $J_\rho$, $q$, $J_q$, $e$,
$J_e$ together give us a closed moment system.
\bibliographystyle{amsplain}

\bibliography{swarm}

\end{document}